# An attack to quantum systems through RF radiation tracking

Kadir Durak , Naser Jam

**ABSTRACT**

A newfound security breach in the physical nature of single photon detectors that are generally used in quantum key distribution is explained, we found that the bit contents of a quantum key transmission system can be intercepted from far away by exploiting the ultrawideband electromagnetic signals radiated from hi-voltage avalanche effect of single photon detectors. It means that in fact any Geiger mode avalanche photodiode that is used inside single photon detectors systematically acts like a downconverter that converts the optical-wavelength photons to radio-wavelength photons that can be intercepted by an antenna as side channel attack. Our experiment showed that the radiated waveforms captured by the antenna can be used as a fingerprint. These finger prints were fed to a deep learning neural network as training data, and after training the neural network was able to clone the bit content of quantum transmission.

## INTRODUCTION

The superexponential time scales of the solutions provided by Quantum computers will likely soon break traditional public key cryptography, including the ciphers protecting most of the world's digital secrets[1-3] . There are many accelerating efforts in development of new Quantum-resistant algorithms for post-quantum cryptography to define the public and private keys but none of them can claim to provide absolute security against future Quantum algorithms. It seems that the only reliable solution on hand at this time that claim absolute security in theory is Quantum cryptography.
Based on the well-known Bohm's version of the Einstein-Podolsky-Rosen experiment [4,5]; the generalized Bell's theorem [6,7] is used to test for eavesdropping. The extraordinary benefit offered by Quantum cryptography is that it is secure against all algorithmic and theoretical advances. For the high security applications, Quantum Key Distribution(QKD) also enables continuous secure communication by use of truly random one-time-pad keys[8].
The cryptographic security of QKD protocols does not rely on any assumptions about the resources available to the adversary and its only problem is that modelling of the implementation in quantum cryptography often enables deviations from an idealized model to be quantified.

## CRYPTOGHRAPHIC SECURITY

Existing fiber based and free space quantum optical links are trusted on the confidentiality of the transmitted bits by the physical properties of single photon transmission.This kind of trust is due to fundamental characteristic of quantum mechanics that any measuring on a quantum system causes disturbance in the system and that a single quantum state cannot be perfectly cloned.[9-14]. In standard quantum cryptographic techniques like E91 and BB84, the transmitter generates secret keys by encoding classical bit values of 0 and 1 using different quantum states of photons in different bases. In the receiver side, photon detectors measure the quantum states of the received photons and converts it to classical bits again. In theory, an eavesdropper disturbs the state of photons and causes bit errors that can be revealed by comparing parts of transmission by cooperation of Sender and Receiver[10,15-16].

Since practical protocols emerged starting in the 1980's and 1990's, QKD has evolved into a thriving experimental field, and is rapidly becoming a solid commercial proposition. In addition to LEO satellite implementations[17], recent deployments with very low noise superconducting single-photon detectors (SPD) that exceeded 421 km[18] range have been reported. However, all theoretical approaches are based on this assumption that every physical element in the system is ideal and free of infirmity. In fact, practical implementations often deviate from the theory, which leaves loopholes for eavesdropping, especially in the physical layer.

## IMPLEMENTATION SECURITY

Extensive studies and experiments over the past years has greatly improved implementation security of QKD. Methods for closing the most readily exploitable loopholes have been developed, but still when the perfect theories of Quantum key transmission are implemented in real world, there are many imperfections.

There are many researches that try to reveal the imperfections of standard protocols regarding the physical limitations like multi-photon emission, weak coherent states, detectors with basis-dependent efficiency, misaligned sources and detectors, and timing jitter of SPDs [20,21]. A summary of typical side-channel attacks against QKD systems including the new technique (subject of this paper) is listed in Table 1 [22].

| Security Issue | Description | Countermeasures |
|---|---|---|
| Trojan-horse attack | Intruder probes the QKD equipment with light to gain information about the device settings | privacy amplification (PA), isolators, filters |
| Multi-photon emission | When more than one photon is emitted in a pulse, information is redundantly encoded on multiple photons | PA, characterization, decoy states, SARG04 and other protocols |
| Imperfect encoding | Initial states do not conform to the protocol | PA, characterization |
| Phase correlation between signal pulses | Non-phase-randomized pulses leak more info to Intruder, decoy states fail | phase randomization, PA |
| Bright-light attack | Intruder manipulates the photon detectors by sending bright-light to them | active monitoring, measurement device independent QKD (MDI-QKD) |
| Efficiency mismatch and time-shift attack | Intruder can control, at least partially, which detector is to click, gaining information on the encoded bit | MDI-QKD, detector summarization |
| Back-flash attack | Intruder can learn which detector clicked and hence knows the bit | isolators, MDI-QKD, detector summarization |
| Manipulation of Local Oscillator reference | In continuous variable QKD (CV-QKD), the local oscillator (LO) can be tampered with by Intruder if it is sent on a communications channel | Generate LO at the receiver. Phase reloading, i.e. only synchronize the phase of LO |
| laser damage attack | creating deviations that leads to side channels by laser-damage | Continuous monitoring of channels |
| RF fingerprint attack (subject of this paper) | Clones single photon detections by using Avalanche electromagnetic pulses | Compact assembly, Shielding and Jamming |

*Table1-List of attacks against a typical QKD system*

## THE NEW MENACE: ELECTROMAGNETIC FINGERPRINT OF AVALANCHE PHOTO DIODES

The most common single-photon detectors used in QKD as shown in Fig. 1-a, are Avalanche Photo Diodes (APDs) operating in reverse bias voltages above the breakdown voltage (Geiger mode), usually over 100Volts. In the Geiger mode the APD becomes so sensitive that detects a single incoming photon. The received photon in detector triggers an avalanche of electrical current, when the current crosses a certain threshold a digital pulse in the output corresponds to photon detection. After that a quenching process sweeps out the avalanche after-currents and prepares the detector for detecting another photon. Direct measurements of discharge current called $I_D(t)$ can show that it has a typical peak current around 10 mA and pulse width of 10 ns with an exponential decay waveform convoluted with a gaussian distribution.

It have been proved experimentally[23], that the charge which is released during the breakdown in a short time produces emissions which generates a fluorescence flash of light.

In addition to light emission during avalanche, we can also expect radiation in RF wavelengths too, because we are facing an accelerating charge in the avalanche process and electrodynamics theories show that we always have far-field electromagnetic radiations when we have an accelerating charge. In the geometrical scales of an APD we can approximate the total charge of the avalanche pulse by a point charge. By integrating the discharge current over the pulse from rise time ($t_r$) to fall time ($t_f$) a total charge $Q_D$ can be calculated by:

$$Q_D = \int_{t_r}^{t_f} I_D(t)dt \qquad (1)$$

By using the non-realistic electrodynamics of an accelerating point charge we can describe radiation by using *Lienard-Wiechert* potential [24], hence by integrating the poynting vector over a sphere the energy radiated per unit time will be :

$$P = \frac{2}{3}\frac{Q_D^2}{c^3}\left(\frac{dv}{dt}\right)^2 \quad for \ \frac{v}{c} \ll 1 \qquad (2)$$

In which $P$ is the total electromagnetic power radiated from the APD, $c$ is speed of light and $dv/dt$ is the acceleration of charge assuming charge speed is much less than $c$.

From a physical point of view we can say that this process is a kind of down conversion occurred in the APD in a form that it converts Optical-wavelength photons to Radio-wavelength photons. From the viewpoint of security this phenomena is a unique feature for each APD because physical location of each APD in surrounding structure and environment is unique and by exploiting the radiation of its RF pulse in the box, shelf, rack, room, etc., the mechanical structure of the environment around the APD acts like a Finite Impulse Response(FIR) filter with unique weights that produces a fingerprint for each APD As shown in Fig. 1-b.

**Figure 1**

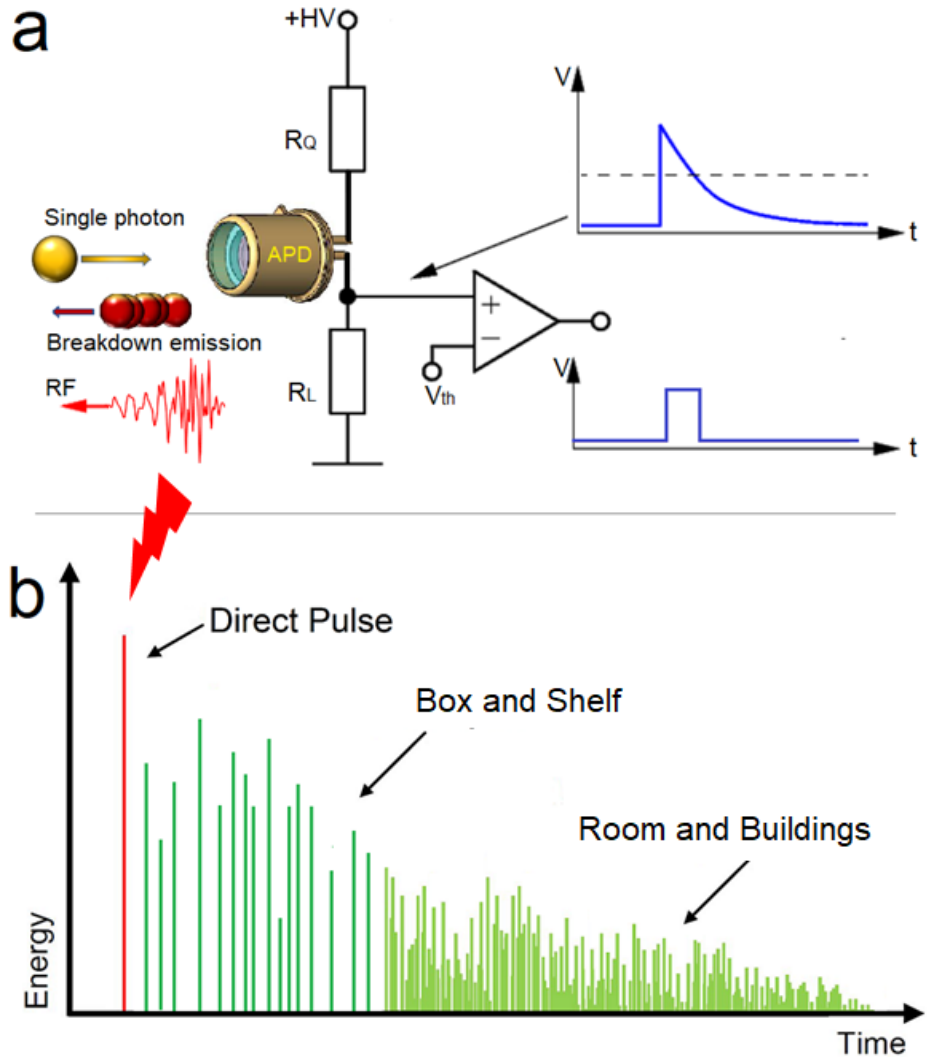

*Avalanche photodiode as a single Photon Detector acts like a downconverter that converts Optical-wavelength single photons to Radio-wavelength photons.(a)Typical circuits of APD for single photon detection, by receiving a single photon hi-voltage avalanche pulse produces RF radiated pulse. (b)Mechanical structure of the system, room and buildings around APD acts like a filter that produces a unique response by multiple reflections for each APD as a fingerprint .*

**A TYPICAL PENETRATION SCENARIO**

Based on mentioned approach many practical scenarios for penetration to a QKD system can be proposed. The proposed approach can be used as a practical Penetration Test (PENTEST) mechanism for ethical hack and evaluating robustness of Quantum cryptography networks that are using single photon based QKD protocols. This approach can be used either in free-space or fiber-optic QKD systems as long as eavesdropper can receive radio waves of the detectors. Also the same concept can be used against the Entanglement-based systems.

One of the popular protocols for QKD is BB84 shown in Fig.2a. In this protocol the idea is to encode every bit of the secret key into the polarization state of a single photon. Because the polarization state of a single photon cannot be measured without destroying this photon, this information will be 'fragile' and not available to the eavesdropper. The protocol then runs in the following steps:

Transmitter (called Alice) sends a sequence of single photon polarized differently. Alice encodes zeroes into H-polarized(0°) photons while unities she encodes into V-polarized photons (90°). But this happens only in half of the cases. The other half of bits, chosen randomly, are encoded using a diagonal polarization basis. Then, the A (45°) polarization corresponds to zero and the D (135°) polarization, to unity. The receiver (called Bob), measures polarization of photons using a detection setup shown in Fig.2a. by this setup receiver can distinguish between H and V polarizations in half of the cases in the '+' basis. But in half of the cases randomly receives photons in another 'x' basis. After transmitting a certain number of bits, Bob announces which basis he used for each bit in a classical communication link. then Alice reveals in which bits they used the same bases. They ignore the bits with different bases, and use only those bits with the same base. After this process the keys are sifted and the length of the key is decreased, but the remaining has randomness and coincidence. Then, they must check eavesdropping by checking the error rate. The test for this criteria is based on Experimental Bell's test [25] based on S which is:

$$S \equiv |\langle \sigma_\alpha \sigma_\beta \rangle + \langle \sigma_\alpha \sigma_{\beta'} \rangle + \langle \sigma_{\alpha'} \sigma_\beta \rangle - \langle \sigma_{\alpha'} \sigma_{\beta'} \rangle| \leq 2 \quad (3)$$

With correlators:

$$\langle \sigma_a \sigma_b \rangle = (1/N_{a,b})(N^{\uparrow\uparrow}{}_{a,b} + N^{\downarrow\downarrow}{}_{a,b} - N^{\uparrow\downarrow}{}_{a,b} - N^{\downarrow\uparrow}{}_{a,b}) \quad (4)$$

Here $N^{AB}{}_{a,b}$ denote the number of events with the respective outcomes A, B for measurement directions a, b and $N_{a,b}$ is the total number of events of the respective measurement setting. Violation of this inequality can be predicted by quantum mechanics when measurements are performed on maximally entangled states:

$$|\Psi_\pm\rangle = (1/\sqrt{2})(|\uparrow\rangle|\downarrow\rangle \pm |\downarrow\rangle|\uparrow\rangle) \quad (5)$$

with certain measurement settings, e.g., α=0°, α′=90°, β=45°, β′=135°.

To this end, Alice and Bob take a part of the key for instance, (10%) and compare it. This procedure is also public, but these 10% are then discarded. If there was an eavesdropping then, the key would contain errors and the whole key is thrown out and the procedure is repeated. The proposed penetration setup for this protocol is shown in Fig.2b andFig.2c. The penetration in performed in the learning and intercept steps as following :

*I-Learning:*

A laser is equipped with a polarization rotator that is synchronized with a precision pulse generator that sends the test photons with random orthogonal polarizations towards the target receiver(Fig. 2b). To achieve a solid relation between received bits and the quantum system, Eve must insert this laser ( that operates in the same wavelength of QKD system) into the link. This laser imitates the Quantum link's laser. Each single photon detector in the target system receives the test photons and generates Hi-voltage avalanche pulses that radiates Ultrawideband (UWB) Electromagnetic pulses (EMP). The EMP generated by avalanche photo-diodes will leak through electromagnetic shields because of high voltage and short pulse-width that produces high bandwidth. This fact have been tested and verified in our experimental setup with two commercial brands of Single Photon Detectors(SPD).

The penetration system is equipped with an ultra-wideband antenna that receives the radiated electromagnetic Pulses. This antenna receives weak avalanche leakages which permits the signal processing section to extract fine differences between pulse fingerprints. The EMP signals are amplified with a low-noise amplifier and excessive noise caused by power lines, mobile communications, etc. is filtered by band-pass and band reject filters. The resulting pure signal is fed to a single-event transient digitizer that is synchronized with the same precision pulse generator that triggers laser pulses.

After preparing the setup, the ultrawideband receive-only antenna with accompanying low-noise amplifier and filter must be located as close to the target photon-detectors as possible.

In this step some test photons are sent to the target by Eve. First random polarizations for transmitted single photons are selected and by receiving RF pulses a histogram of polarizations is formed which the peak of histogram shows the right polarization. After that random locations for the antenna is selected and locations are improved incrementally by a successive approximation algorithm. In the signal processor successive integration of the test pulses are performed to achieve highest possible signal-to-noise ratio (SNR).Strong and discriminated waveforms captured by the signal processor are stored in the memory as a reference for machine learning training. In this stage Alice and Bob know that eavesdropping took place because the keys would contain excessive errors. Then the whole key is thrown out and the BB84 procedure is repeated again by Alice and Bob.

*II-Key Intercept:*

In this step of the operation(Fig 2c), Eve's laser is removed and the target QKD system operates in its normal condition and the Eve starts to capture avalanche signals of the SPDs. The signal processor and the artificial intelligence engine recognizes the received signals in a free-running form without any trigger signal. By this configuration Eve doesn't need to make any interference in QKD path and her only source of information is the classical communication path and the SPD's RF backdoor, hence from the viewpoint of information theory in ideal form (which the Artificial Intelligence engine can perfectly identify different SPD waveforms without error) Eve has access to the same information of Bob. It is worth mentioning that the techniques proposed for waveform detection in this setup are mature techniques that are used normally in Ultrawideband Radars and electronic warfare systems [26-31].

# Figure 2

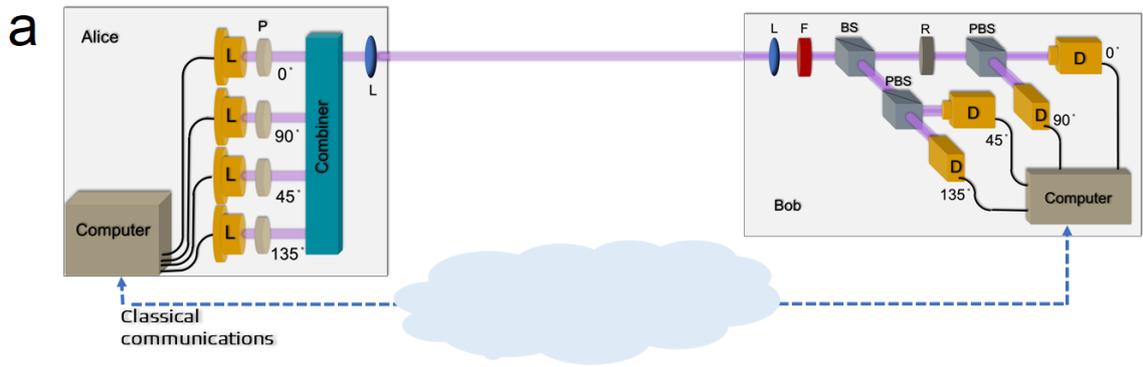

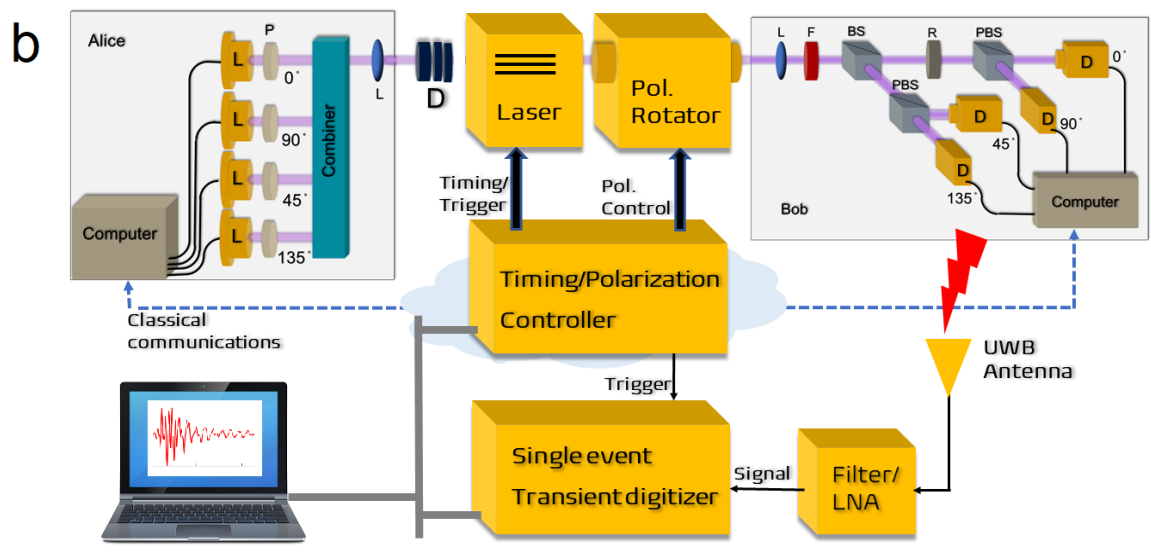

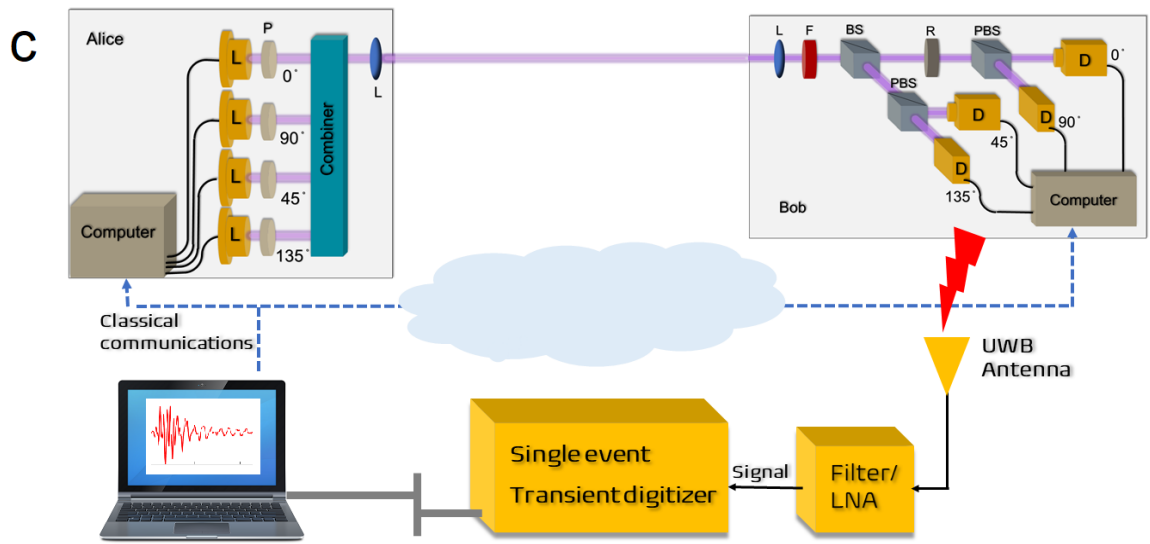

Penetration setup for a Quantum Key Distribution system: (a)Typical BB84 QKD system, (b)Penetration Training phase, (c)Eavesdropping phase .

# EXPERIMENTAL SETUP

To evaluate the possibility of automatic differentiation between fingerprint of radio waves that are received from different SPD locations we prepared a setup shown in Fig. 3 according to ITU standards [32]. The setup was including a digital Scope as transient digitizer, an Ultrawideband antenna and two commercial single photon detectors with same model/manufacturer.

After setting up the components, 64 waveforms were captured from each SPD with 1GS/s sampling rate and 1200 samples per waveform. Fig. 4 shows one of the captured raw waveforms corresponding to a synchronous digital pulse created by SPD. The TTL output pulse of the APD is used as a synchronization pulse for precise data acquisition from the antenna.

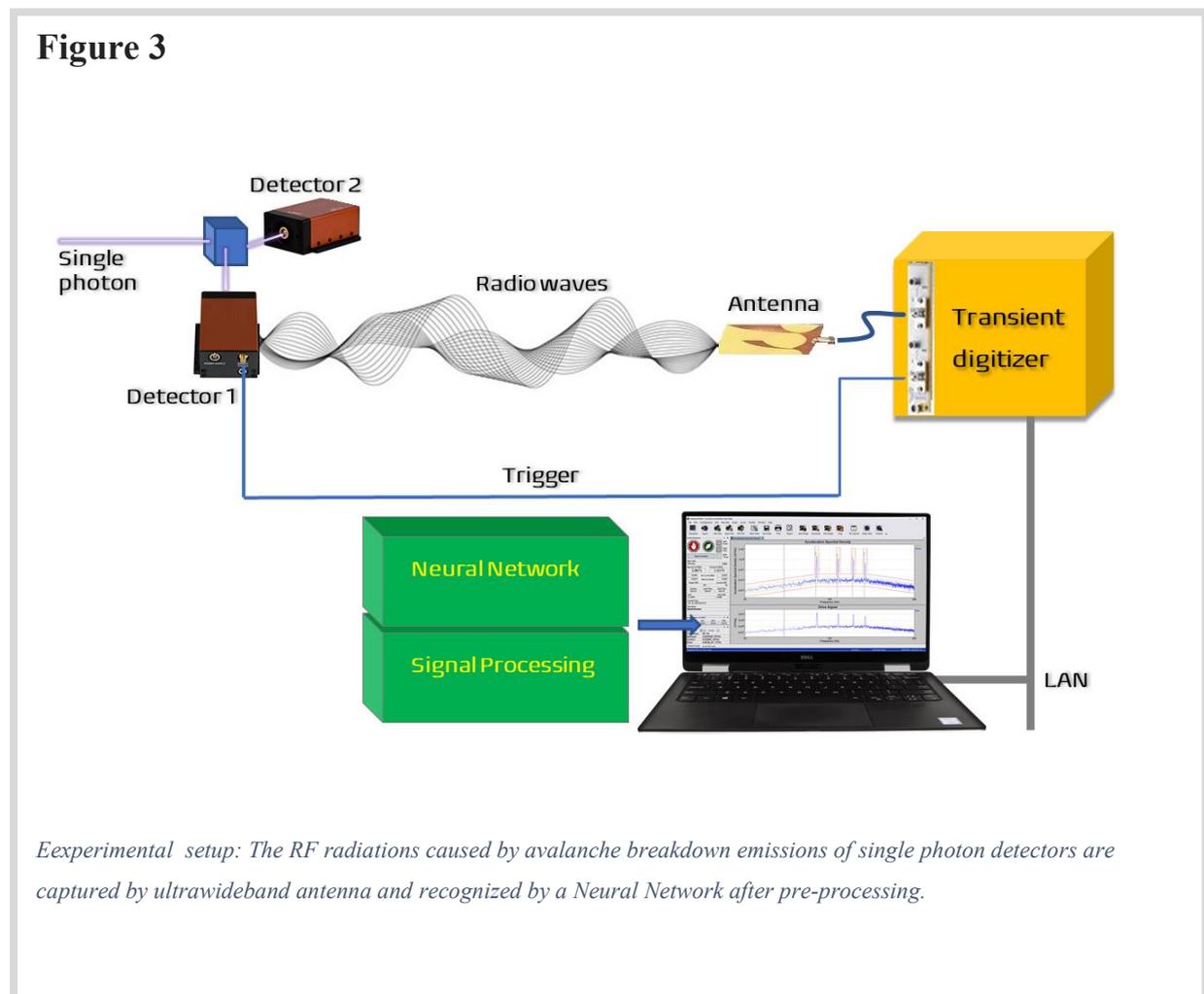

**Figure 3**

*Eexperimental setup: The RF radiations caused by avalanche breakdown emissions of single photon detectors are captured by ultrawideband antenna and recognized by a Neural Network after pre-processing.*

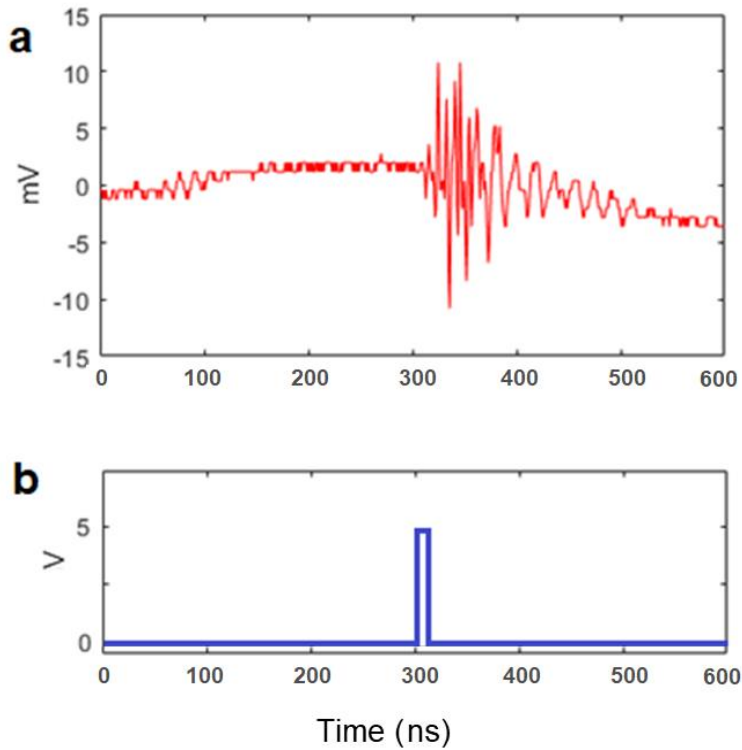

*Real ultrawideband Electromagnetic pule produced by Hi-voltage avalanche impulse of APD: (a)Raw waveform captured by the ultrawideband antenna, (b)TTL output pulse of the APD used as trigger for signal acquisition.*

## SIGNAL PROCESSING

In the signal processing procedure, we removed the noises and irrelevant portions of the signal by frequency domain excision (filtering ) and time domain excision. The frequency domain excision shown in Fig. 5 removes both low-frequency noises induced by sources like powerline and high frequency noises that appear in the background when no signal is present. After removing noises by frequency domain excision, the signal samples were decreased from 1200 samples to 256 samples per waveform by time domain excision, in a way that noises before and after the signals of interest were cut out from the signals. This operation not only increases the signal-to-noise ratio but also removes the effect of irrelevant parts of the signal that contains no information from further processing. To perform a correlation analysis we made an additional processing to normalize each waveform. This normalization was performed by dividing each sample to the total power of the waveform.

**Figure 5**

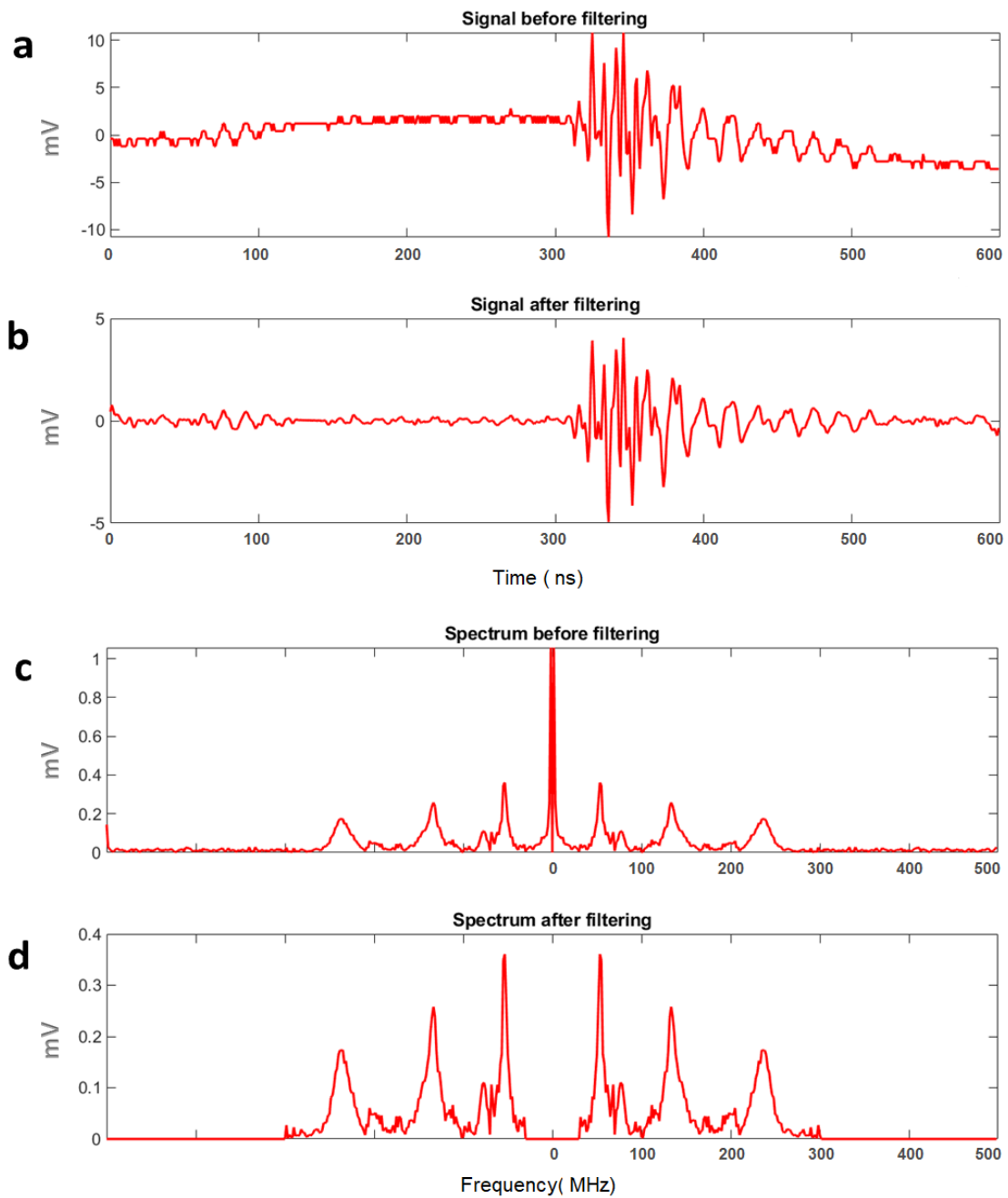

*Noise removal by frequency domain excision: (a)Signal before filtering, (b) Signal after filtering, (c)Spectrum before filtering, (d)Spectrum after filtering.*

Before any further processing, to obtain an overview about the amount of decorrelation of the waveforms captured from two separate SPDs, we picked a waveform from first SPD as a reference. Then we calculated the absolute value of cross-correlations of this reference with 64 waveforms of the same SPD and 64 waveforms of the second SPD, the results are shown in Fig. 6. Here we put the correlation results of the SPD1 (named co-location) and SPD2 (named cross-location) beside each other on a common scale to have an overview and physical sense about the amount of decorrelation. This decorrelation can be described as a fingerprint for each SPD by this physical property that impulses from each Avalanche photodetector experience different paths and reflections from the physical objects around (mounting boxes, shelf, rack, walls of the room, …) that act like systems having different impulse responses, let's call them $h_1[n]$ and $h_2[n]$ in discrete-time form. Then if we recall that the impulse produced by each Avalanche photodiode is similar to a Dirac delta function that we call it $\delta[n]$ then if we take 64 waveforms from SPD1 and 64 waveforms from SPD2 we should have the following waveforms:

$$W_1[n, i] = h_1[n] * \delta_1[n, i] + N[n, i] + Q[n, i] \quad ; i=1 \text{ to } 64 \quad (6)$$

$$W_2[m, i] = h_2[m] * \delta_2[m, i] + N[m, i] + Q[m, i] \quad ; i=1 \text{ to } 64 \quad (7)$$

In which, $W_1[n, i]$ is the i-th waveform received by the antenna from SPD1, $h_1[n]$ is the impulse response of the path from SPD1 to the antenna, $\delta_1[n, i]$ the i-th impulse of the Avalanche photodiode of SPD1 after detecting a single photon and $N[n, i]$ is the additive white Gaussian thermal noise of the receiving system and $Q[n, i]$ is the quantization error of the transient digitizer which is modelled as an additive random noise.

Here $m = n + \Delta t$ in which $\Delta t$ is caused by time difference of arrival of waveforms to antenna due to different distances of each SPD to antenna.

There are three sources that construct each waveform: the SPD, the radiation path, and the receiving system. The main player that has the first role in making decorrelation between SPD waveforms is the radiation path between two SPDs. As it can be seen in the captured signal's spectrum there are three peaks in the FFT output that may correspond to resonance caused by the cavity of SPD box and shelf. The shelf was used to make the test scenario more realistic.

**Figure 6**

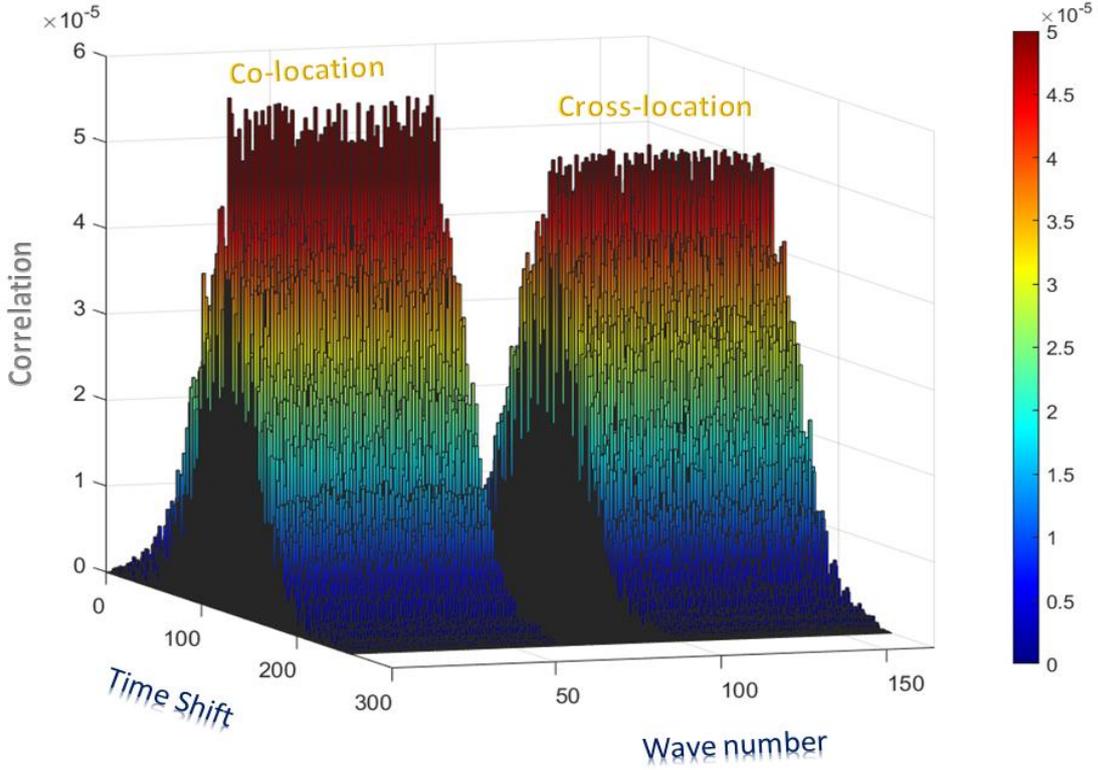

*Cross correlations of 64 waveforms captured from the SPDs in one place (Co-location) and two different places (Cross-location) with 20 cm distance.*

After obtaining an overview about the amount of decorrelation of waveforms, to make sure that there is enough decorrelation between fingerprints of two SPDs we calculated matrices of co-location peak values and cross-location correlations of the waveforms. By this way we obtain two matrices as following:

$$M_{co}[i,j] = peak\{|R_{ij}[\tau]|\} \; ; \; i,j = 1 \text{ to } 64 \quad (8)$$

$$M_{cross}[i,j] = peak\{|R_{ij}[\tau]|\} \; ; \; i,j = 1 \text{ to } 64 \quad (9)$$

In which, $R_{ij}[\tau]$ is cross correlation of i-th and j-th waveforms and $M_{co}$ is co-location matrix and $M_{cross}$ is cross-location matrix. To have a thorough view that makes sense in proof-of-concept we made two 3-dimentional heat-map graphs that shows $M_{co}$ and $M_{cross}$ correlation matrices in the same scale. The graphs are shown in Fig. 7.

As it can be seen in the heatmap graphs, in our proof-of-concept prototype, two SPDs were mounted 20 cm apart and there was enough decorrelation between two SPD locations, hence the correlation matrixes can be separated by a flat surface as a threshold. It means that an artificial intelligence algorithm may be used to learn and differentiate between the waveforms of each APD in a fast and automated way. A survey conducted on 2019[33] shows that deep learning techniques can be effectively used in different wireless signal recognition scenarios and applying the raw signal to the Neural network without any preprocessing is becoming a trend in signal recognition.

## SIGNAL RECOGNITION USING DEEP LEARNING NEURAL NETWORK

In this stage we implemented a deep learning signal classification Neural network to differentiate between two SPD waveforms. In fact, this machine automates the recognition of decorrelations between waveforms after signal processing and filtering. Fig. 8 shows a simplified view of the configuration of the deep learning Neural network used for training and waveform recognition.

In our proof-of-concept setup we used half of the waveforms for training purpose and half for test. Our Neural network's number of input neurons was equal to the number of samples of input signal. after time-excision, each input neuron was containing a time-sample of a waveform, then we made hidden layers of fully interconnected neurons and finally prepared one output neuron with binary output which is trained to deliver two values representing the received waveforms of SPDs in which 'zero' corresponds to SPD1 and 'one' corresponds to SPD2. After training we applied the test waveforms to the network and the results showed that the network can classify the waveforms received from each SPD with an accuracy of better than 99.7 percent. It shows that we can clone most of the single photon bits received in Bob side with very low error rate and violate the quantum safety of the QKD link.

**Figure 7**

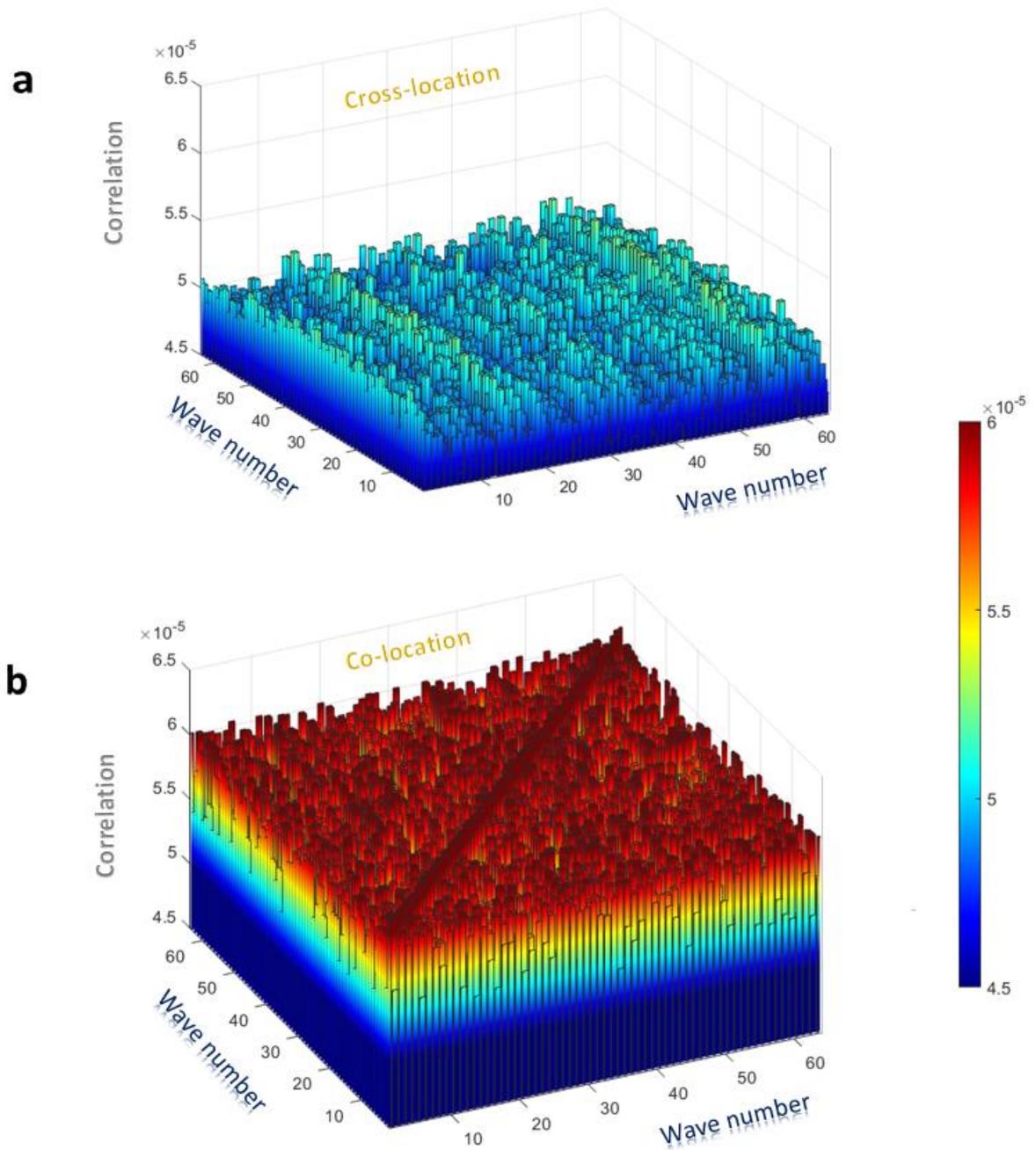

*Cross-correlation matrices made by 64 waveforms captured in one place(Co-location) and two different places(Co-location) : (a) Cross-correlation matrix of 64 waveforms captured from SPDs in different locations. (b) Correlation matrix of 64 waveforms captured from an SPD in one location.*

**Figure 8**

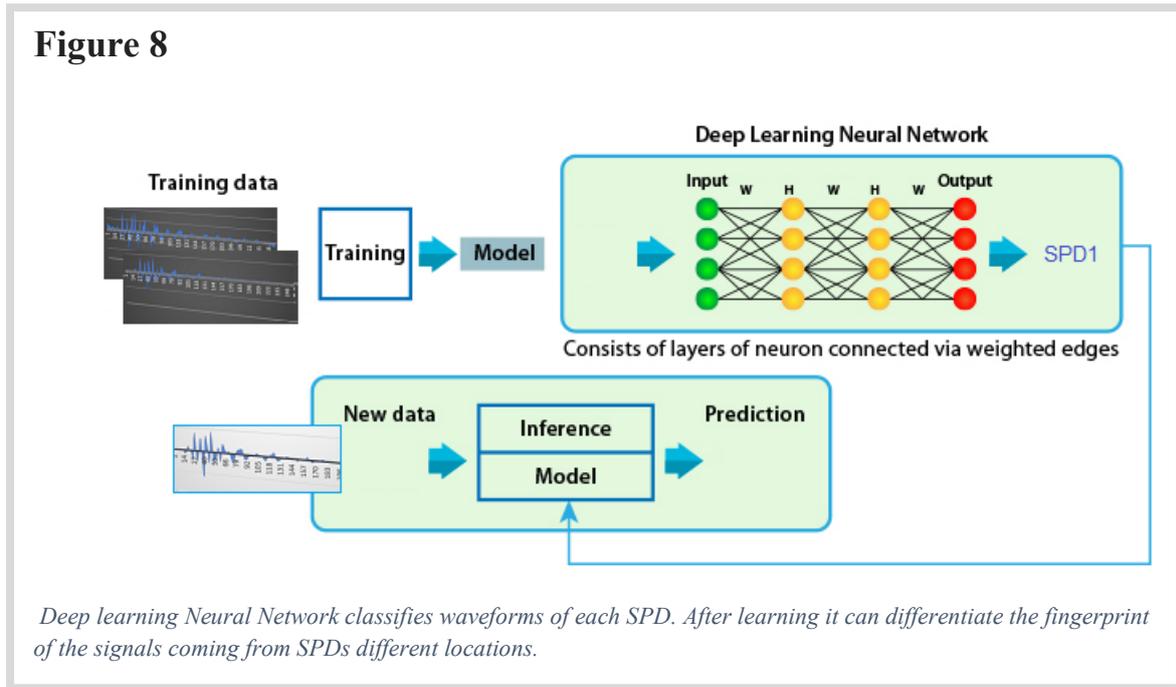

*Deep learning Neural Network classifies waveforms of each SPD. After learning it can differentiate the fingerprint of the signals coming from SPDs different locations.*

## DISCUSSION

Experimental tests on commercial SPDs showed that the ultrawideband high-voltage EMP of avalanche photodiodes can penetrate through the shielding of SPDs and can be intercepted by an eavesdropper without touching the error rate of the quantum link . In this experiment we have shown that it is possible to capture and differentiate between impulses received from different SPDs by propagation of fingerprint of signals. During the tests it have been observed that fingerprint decorrelations are highly dependent on the spacing between two SPDs. Reduction of spacing not only increases the pattern-recognition error rate of the Eavesdropper but also increases the thermal and quantization noise of the transient digitizer because of need for more bandwidth and sampling speed. This causes more error in the Eve's side than Bob's. It can be said that although this experiment is not the end of the matter for Quantum key distribution but it shows that any secure communication system that relies on absolute safety of quantum transmission must also include relevant Cyber-physical and Electronic warfare techniques (like high shielding and Jamming), to ensure that any kind of hostile use of electromagnetic spectrum (not only the light wavelengths) should be seriously taken into consideration.

As a conclusion we can say that our experiment showed that at least the following provisions should be seriously prepared for every Quantum-secured communication system:

1- *Assembling the SPDs as close to each other as possible*
2- *Maximum Electromagnetic shielding  utilization in physical layer*
3- *Including a wideband Electromagnetic jammer inside the Quantum receiver system*

# METHODS

The feasibility study of this project was performed using a home-made, passively quenched single photon detector based on Geiger APDs. We used two types of commercial ADP unites with 500μm diameter active area and hermetically sealed metal package, the first one was type SAP500 from 'Laser components' and the second one C30902SH-TC from 'Excelitas Technologies'. After positive results of the feasibility study, in the next level of experiments we used a commercial single photon detector type ID100 from ' ID Quantique SA' company that is an active and famous company in the Quantum security business with mature professional QKD products.

Signal reception was performed using a home-made wideband antenna with 0dB gain and a wideband digital oscilloscope with 1GS/s sample rate. The distance between two SPDs were 20 cm and distance between SPD setup and receiving antenna was 2 meters . In order to eliminate any uncertainty that may cause by moving objects, the transient recorder was remotely controlled via LAN connection. All signal processing and transforms was performed using MATLAB's signal processing toolbox. Signal Bandpass filtering was performed by FFT excision with cut-off frequency of 30 MHz to 300 MHz . The Neural network training and tests was performed using MATLAB's Deep learning toolbox. We implemented a deep learning Neural network with 256 inputs, each containing a time-sample of a waveform , then we made 5 layers of fully interconnected hidden neurons and one binary output neuron that presents the bit information of quantum transmission.

# DATA AVAILABILITY

The data that support the plots within this paper and other findings of this study are available from the corresponding author upon reasonable request.

# AFFILIATIONS

Dr. Kadir Durak is head of Quantum optics lab in Ozyegin university at Istanbul & Naser Jam is a member of Quantum optics lab in Ozyegin university.